\documentstyle[aps,psfig,multicol]{revtex}
\newcommand{\bcols}{\ifpreprintsty\else\begin{multicols}{2}\fi}
\newcommand{\ecols}{\ifpreprintsty\else\end{multicols}\fi}
\begin{document} 
\draft
\title{Quantum-optical communication rates through an
amplifying random medium} 
\author{J. Tworzyd\l o$^{\rm a,b}$ and C. W. J. Beenakker$^{\rm a}$} 
\address{$^{\rm a}$Instituut-Lorentz, Universiteit Leiden,
P.O. Box 9506, 2300 RA Leiden, The Netherlands\\
$^{\rm b}$Institute of Theoretical Physics, Warsaw University,
Hoza 69, 00-681 Warszawa, Poland} 
\date{March 19, 2002}
\maketitle 
\begin{abstract}
We study the competing effects of stimulated and spontaneous emission on the
information capacity of an amplifying disordered waveguide. At the laser
threshold the capacity reaches a ``universal'' limit, independent of the
degree of disorder. Whether or not this limit is larger or smaller than the
capacity without amplification depends on the disorder, as well as on the 
input power. Explicit expressions are obtained for heterodyne detection of
coherent states, and generalized for arbitrary detection scheme.
\end{abstract} 
\pacs{PACS numbers: 89.70.+c, 42.50.Ar, 42.50.Lc, 42.25.Dd }

\bcols
To faithfully transmit information through a communication channel, the rate of
transmission should be less than the capacity of the channel
\cite{Sha48,Cov91}. Although current technology is still far from the
quantum limit, there is an active scientific interest in the fundamental
limitations to the capacity imposed by quantum mechanics \cite{Cav94,Nie00}.
Ultimately, these limitations originate from the uncertainty principle, which
is the source of noise that remains when all external sources have been
eliminated.

An important line of investigation deals with strategies to increase the
capacity. One remarkable finding of recent years has been the beneficial role
of multiple scattering by disorder, which under some circumstances can increase
the capacity by increasing the number of modes (scattering channels) that
effectively carry the information \cite{Fos96,Sim01}. Quite generally, the
capacity increases with increasing signal-to-noise ratio, so that amplification
of the signal is a practical way to increase the capacity. When considering the
quantum limits, however, one should include not only the amplification of the
signal (e.g.\ by stimulated emission), but also the excess noise (e.g.\ due to
spontaneous emission). The two are linked at a fundamental level by the
fluctuation-dissipation theorem, which constrains the beneficial
effect of amplification on the capacity \cite{Cav82}.

While the effects of disorder and amplification on communication rates have
been considered separately in the past, their combined effects are still an
open problem. Even the basic question ``does the capacity go up or down with
increasing gain?'' has not been answered. We were motivated to look into this
problem by the recent interest in so called ``random lasers''
\cite{Wie97,Cao01}. These are optical media with gain, in which the feedback is
provided by disorder instead of by mirrors (as it is in a conventional laser).
Below the laser threshold these materials behave like linear amplifiers with
strong intermode scattering, and this results in some unusual noise properties
\cite{Bee98,Pat99}. As we will show here, the techniques developed in connection with
random lasers can be used to predict under what circumstances the capacity is
increased by amplification.

\begin{figure}
\centerline{\psfig{figure=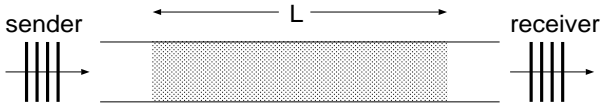,width=8cm}\medskip}
\caption[]
{Communication channel 
consisting of an $N$-mode waveguide 
that is amplifying over a 
length $L$. Both sender and
receiver use a single narrow-band mode (indicated by
a plane wave).}

\label{system}
\end{figure}

We consider the transmission of information through a
linear amplifier consisting of an $N$-mode
waveguide that is pumped uniformly over a length $L$
(see Fig.\ \ref{system}).
We will refer to amplification by stimulated emission,
but one can equally well assume other gain mechanisms (for example,
stimulated Raman scattering \cite{Man95}). 
The amplification occurs at a rate
$1/\tau_a$. The waveguide also contains passive scatterers,
with a transport mean free path $l$. The combined effects of 
scattering and amplification are described by a $2N\times2N$ 
scattering matrix $S$ which is super-unitary 
($SS^{\dagger}-\openone$ positive definite).

We assume that the information itself is of a classical nature 
(without entanglement of subsequent inputs), but fully account for 
the quantum nature of the electromagnetic field that carries 
the information. The quantized radiation is described by a 
vector $a^{\rm in}$ of bosonic annihilation operators for the 
incoming modes and a vector $a^{\rm out}$ for the outgoing 
modes. The two vectors are related by the input-output
relation \cite{Bee98,Jef93,Gru96}
\begin{equation}
a^{\rm out} = S a^{\rm in} + U b^{\dagger}.
\label{in_out}
\end{equation}
The vector of bosonic creation operators $b^{\dagger}$
describes spontaneous emission by the amplifying medium.
The fluctuation-dissipation theorem relates the matrix $U$ to $S$
by
\begin{equation}
UU^{\dagger}=SS^{\dagger}-\openone.
\label{fluct_diss}
\end{equation}

The first communication channel that we examine 
is heterodyne detection
of coherent states \cite{Cav94}. The sender uses a single narrow-band
mode $\alpha$ (with frequency $\omega_0$ and band width $\Delta \omega$),
to transmit a complex number 
$\mu$ by means of a coherent state $|\mu\rangle$ (such that 
$a^{\rm in}_\alpha|\mu\rangle=\mu|\mu\rangle$). The receiver measures a
complex number $\nu$ by
means of heterodyne detection of mode $\beta$. Two sources of noise
may cause $\nu$ to differ from $\mu$: Non-orthogonality
of the two coherent states $|\mu\rangle$ and $|\nu\rangle$; and spontaneous
emission by the amplifying medium.

The {\it a priori} probability $p(\mu)$ that the sender transmits
the number $\mu$, and the conditional probability
${\mathcal P}(\nu|\mu)$ that the receiver detects 
$\nu$ if $\mu$ is transmitted, 
determine the mutual information \cite{Cav94}
\begin{equation}
I = \int\, {\mathrm d}^2\nu \, \int \, {\mathrm d}^2\mu\, 
{\mathcal P}(\nu|\mu) p(\mu) \log_2\left(\frac{{\mathcal P}(\nu|\mu)}{\tilde{p}(\nu)}\right).
\label{mutual}
\end{equation}
We have defined 
$\tilde{p}(\nu)=\int\, \mathrm{d}^2\mu\, {\mathcal P}(\nu|\mu) {\mathit p}(\mu)$.
The channel capacity $C$ (measured in bits per use of the channel)
is obtained by maximizing $I$ over the {\it a priori} distribution $p(\mu)$,
under the constraint of fixed input power  $P= P_0 \int\, {\mathrm d}^2\mu\,
|\mu|^2 p(\mu)$ (with $P_0=\hbar \omega_0 \Delta\omega/2\pi$).
As argued in Ref.\cite{Mou00}, any randomness in the scattering medium
that is known to the receiver but not to the sender can be 
incorporated by averaging $I$ before maximizing, hence
\begin{equation}
C= \max_{p(\mu)} \left< I \right>.
\label{maxima}
\end{equation}
The brackets $\left< \cdots \right>$ indicate an average
over different positions of the scatterers.

The calculation of the capacity is greatly simplified 
by the fact that the spontaneous emission noise is
a Gaussian superposition of coherent states. This is
expressed by the density matrix of the amplifying medium,
\begin{equation}
\rho_{\rm medium}\propto \int \, {\mathrm d}^2\vec{\beta}\, 
\exp(-|\vec{\beta}|^2/f) 
|\vec{\beta}\rangle\langle\vec{\beta}|,
\label{rho_medium}
\end{equation}
where $\vec{\beta}$ is a vector of $2N$ complex
numbers and $|\vec{\beta}\rangle$ is the corresponding coherent 
state (such that $b_n|\vec{\beta}\rangle=\beta_n|\vec{\beta}\rangle$).
The variance $f= N_{\rm upper} (N_{\rm upper}-N_{\rm lower})^{-1}$
depends on the degree of population inversion of the upper
and lower atomic levels
that generate the stimulated emission. Minimal noise requires a complete
population inversion: $N_{\rm lower}=0$ $\Rightarrow$ $f=1$. We
will consider that case in what follows.

We will similarly assume that the heterodyne detection
adds the minimal amount of noise to the signal.
(This requires that the image band is in the vacuum 
state \cite{Cav94}.) The conditional probability  is then given simply
by a projection,
\begin{equation}
{\mathcal P}(\nu|\mu)=\langle\nu|\rho_{\rm out}(\mu)|\nu\rangle,
\label{cond_prob}
\end{equation}
of the density matrix $\rho_{\rm out}(\mu)$ of the outgoing mode $\beta$
onto the coherent state $|\nu\rangle$ (for an incoming coherent state
$|\mu\rangle$ in mode $\alpha$). In view of Eqs.\ (\ref{in_out}) and 
(\ref{rho_medium}) we have (for $f=1$)
\begin{equation}
\rho_{\rm out}(\mu) \propto \int\, d^2\nu'\,
\exp\left( 
- \frac{ |\nu'-S_{\beta \alpha}\mu|^2}{\sum_n|U_{\beta n}|^2}
\right)
|\nu'\rangle\langle\nu'|.
\label{rhoout}
\end{equation}
This is again a Gaussian superposition of coherent states, but
now the variance is related by Eq.\ (\ref{fluct_diss}) to the scattering 
matrix of the medium: 
$\sum_n |U_{\beta n}|^2=\sum_n |S_{\beta n}|^2-1$.

Substituting $\rho_{\rm out}$ into Eq.\ (\ref{cond_prob}), and using
$|\langle\nu|\nu'\rangle|^2=\pi^{-1}\exp(-|\nu-\nu'|^2)$, we arrive at
\begin{equation}
{\mathcal P}(\nu|\mu)\propto \exp\left(
-\frac{|\nu-S_{\beta \alpha}\mu|^2}{\sum_n|S_{\beta n}|^2} 
\right).
\end{equation}
This expression for the conditional probability has the same Gaussian
form as in previous studies \cite{Mou00,Hal94} of communication channels 
degraded by Gaussian noise, but the essential difference is
that in our case the noise strength is not independent
of the transmitted power, but related to it by the 
fluctuation-dissipation theorem (\ref{fluct_diss}). 

The calculation of the capacity proceeds as in 
Refs.\cite{Mou00,Hal94}. The optimum {\it a priori} distribution 
$p(\mu) \propto \exp(-|\mu|^2 P_0/P)$
is independent of the
scattering matrix $S$, so the maximization and disorder 
average in Eq.\ (\ref{maxima}) may be interchanged. The result is
\begin{equation}
C = \left\langle \log_2 \left(
1+ {\cal R}  
\right)\right\rangle,\;\;  
{\cal R}= \frac{(P/P_0) |S_{\beta \alpha}|^2}
{\sum_{n=1}^{2N} |S_{\beta n}|^2}.
\label{capacity}
\end{equation}
The quantity ${\cal R}$ is the signal-to-noise 
ratio at the receiver's end.
We can write ${\cal R}$ equivalently in terms of the
transmission matrix $t$ (from sender to receiver) and the
reflection matrix $r$ (from receiver to receiver):
\begin{equation}
{\cal R} = \frac{(P/P_0) |t_{\beta\alpha}|^2}
{\sum_{n=1}^N(|t_{\beta n}|^2+|r_{\beta n}|^2)}.
\end{equation}

In the absence of intermode scattering one has 
$|t_{n m}|^2 = \delta_{nm}$ and $r_{nm}=0$, 
hence ${\cal R}= \delta_{\alpha\beta}P/P_0$ and 
$C=\log_2(1+\delta_{\alpha\beta}P/P_0)$, 
independent of the amount of amplification.
The increase in capacity by
stimulated emission is canceled by the extra noise from 
spontaneous emission \cite{Cav82}. In the absence of amplification, 
but in the presence of scattering, one has $\sum_n|S_{\beta n}|^2 = 1$,
hence $C=\langle\log_2(1+|t_{\beta\alpha}|^2 P/P_0)\rangle$. 
The capacity is 
reduced by inter-mode scattering in the same way as for the lossy channel 
studied in Ref.\cite{Ari98}.

\begin{figure}
\centerline{\psfig{figure=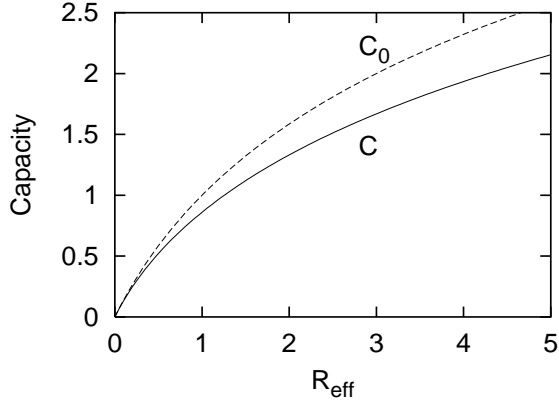,width=8cm}\medskip}
\caption[]
{Capacity $C$ for heterodyne detection of coherent states
as a function of the
signal-to-noise ratio ${\cal R}_{\rm eff}$.
The result (\ref{result}) lies below the value 
$C_0=\log_2(1+{\cal R}_{\rm eff})$
that ignores statistical fluctuations. }
\label{plot}
\end{figure}

The average over the scatterers
can be done analytically in the limit $N\gg 1$ of a 
large number of modes in the waveguide. Sample-to-sample fluctuations
in the denominator $\sigma = {\sum_n(|t_{\beta n}|^2+|r_{\beta n}|^2)}$
are smaller than
the average by an order $N$, so these fluctuations may be 
neglected and we can replace the denominator by its average
$\bar{\sigma}$.
The fluctuations in the numerator $\tau=|t_{\beta\alpha}|^2$
can not be ignored. These are described (for $N\gg 1$) by the
Rayleigh distribution
${\cal P}(\tau)= \bar{\tau}^{-1} e^{- \tau/\bar{\tau}}$.
Integrating $\log_2(1+(P/P_0)\tau/\bar{\sigma})$ over $\tau$
with distribution ${\cal P}(\tau)$ we arrive at
\begin{equation}
C= e^{ {\cal R}_{\rm eff}^{-1} } \Gamma(0;{\cal R}_{\rm eff}^{-1})/\ln 2,
\;\;
{\cal R}_{\rm eff} =  \frac{P \bar{\tau}}{P_0 \bar{\sigma}},
\label{result}
\end{equation}
with $\Gamma(0;x)$ the incomplete gamma function. The
dependence of the capacity $C$ on the effective signal-to-noise
ratio ${\cal R}_{\rm eff}$ is plotted in Fig.\ \ref{plot}.
It lies always below the capacity $C_0=\log_2(1+{\cal R}_{\rm eff})$
which one would obtain by ignoring fluctuations in $\tau$.
For ${\cal R}_{\rm eff} \ll 1$ the two capacities approach
each other, $C \approx C_0 \approx {\cal R}_{\rm eff}/\ln 2$,
while for ${\cal R}_{\rm eff} \gg 1$ one has 
$C_0 \approx \log_2{\cal R}_{\rm eff}$ versus
$C \approx \log_2{\cal R}_{\rm eff}-\gamma/\ln 2$
(with $\gamma \approx 0.58$ Euler's constant).

The quantity ${\cal R}_{\rm eff}$ depends on three length
scales, the length $L$ of the amplifying region, the
mean free path $l$, and the amplification 
length $l_a=\sqrt{D \tau_a}$ (with $D$ the diffusion
constant).
The two averages $\bar{\tau}$, $\bar{\sigma}$ 
can be calculated from the diffusion equation in the regime
$l\ll l_a,L$. There is a weak dependence on the mode indices
$\alpha, \beta$ in this diffusive regime, which we ignore
for simplicity. The result is
\begin{equation}
\bar{\tau} = \frac{4 l/3 l_a}{N \sin(L/l_a)},\;\;
\bar{\sigma} = 1+ (4 l/3 l_a)\frac{1-\cos(L/l_a)}{\sin(L/l_a)}.
\end{equation}
The effective signal-to-noise ratio becomes
\begin{equation}
{\cal R}_{\rm eff} = \frac{P}{N P_0} \left[
1-\cos(L/l_a)+(3l_a/4l)\sin(L/l_a)
\right]^{-1}.
\label{reff}
\end{equation}
Without amplification, for $l_a\gg L$, one has
${\cal R}_{\rm eff}=\frac{4}{3} (l/NL) P/P_0$.
Amplification increases ${\cal R}_{\rm eff}$, up
to the limit ${\cal R}_{\rm eff}\rightarrow P/2NP_0$ that
is reached upon approaching the laser threshold $l_a\rightarrow L/\pi$.

We conclude that amplification in a disordered waveguide increases
the capacity for heterodyne detection of coherent states, up to the limit
\begin{equation}
C_{\infty}= e^{2NP_0/P} \Gamma(0;2NP_0/P)/\ln 2
\label{infty}
\end{equation}
at the laser threshold. This limit is ``universal'', in the
sense that it is independent of the degree of disorder (as long
as we remain in the diffusive regime). We have 
$C_{\infty}\approx P/2 NP_0 \ln 2$ for $P\ll NP_0$ and 
$C_{\infty}\approx \log_2(P/2 NP_0)-\gamma/\ln 2$ for $P\gg NP_0$. 
The increase in the capacity
by amplification in the diffusive regime is therefore up to
a factor $3L/8l$ for $P\ll NP_0$ and up to 
a factor $1+(\ln L/l)(\ln P/N P_0)^{-1}$ for $P\gg NP_0 (L/l)$.
All this is in contrast to the case of a waveguide without disorder,
where the capacity is independent of the amplification.

We now relax the 
requirement of heterodyne detection and instead consider
the maximum communication  rate for any physically
possible detection scheme \cite{Cav94}. We still assume
that the information is encoded in coherent states, and
use the same Gaussian {\it a priori} distribution
$p(\mu) \propto \exp(-|\mu|^2 P_0/P)$ as before.
It has been conjectured \cite{Hol01} that 
an input of coherent states with this Gaussian distribution
actually maximizes the information rate for any method of
non-entangled input with a fixed mean power (the so called 
one-shot unassisted classical capacity).

The capacity for arbitrary detection scheme is given by the 
Holevo formula \cite{Sch97,Hol98a}
\[
C_{\rm H} = H\left(\int\, {\mathrm d}^2\mu \, p(\mu) \rho_{\rm out}(\mu)\right) - 
\int\, {\mathrm d}^2\mu \, p(\mu) H\left[\rho_{\rm out}(\mu)\right],
\]
where $H(\rho)=-{\rm Tr}\rho\log_2\rho$
is the von Neumann entropy.
For a Gaussian density matrix 
$\rho \propto \int\, {\mathrm d}^2\mu \, \exp(-|\mu-\mu_0|^2/x)$
one has \cite{Hol99}
\begin{equation}
H(\rho) =(x+1)\log_2(x+1)-x\log_2 x \equiv g(x).
\end{equation}
Applying this formula to the
Gaussian $\rho_{\rm out}(\mu)$  in Eq.\ (\ref{rhoout}) 
and the Gaussian $p(\mu)$, we arrive at the capacity 
\begin{equation}
C_{\rm H} = g(\tau P/P_0+\sigma-1)-g(\sigma-1).
\end{equation}

For a channel without amplification $\sigma\rightarrow 1$ and
so $C_{\rm H}=g(\tau P/P_0)$, which lies above the capacity for
heterodyne detection considered earlier. At the other extreme,
upon approaching the laser threshold, $\sigma\rightarrow \infty$
and we have $C_{\rm H} \rightarrow \log_2(\tau P/\sigma P_0)$, which is
the same limiting expression as for heterodyne detection.

The average over disorder can be carried out as before by replacing
$\sigma$ by $\bar{\sigma}$ and averaging over $\tau$ with
the Rayleigh distribution ${\cal P}(\tau)$. The result is
\begin{eqnarray}
C_{\rm H} = \frac{\bar{\tau}P}{P_0} \log_2\frac{\bar{\sigma}}{\bar{\sigma}-1}
+ \frac{\bar{\tau}P}{P_0\ln 2} &&
\bigl[ e^{{\cal R}_{\rm eff}^{-1} } \Gamma(0;{\cal R}_{\rm eff}^{-1}) \nonumber \\
&& - e^{{\cal R}_{\rm eff}'^{-1} } \Gamma(0;{\cal R}_{\rm eff}'^{-1})
\bigr],
\label{averchi}
\end{eqnarray}
where ${\cal R}_{\rm eff}/{\cal R}_{\rm eff}'=1-1/\bar{\sigma}$.

\begin{figure}
\centerline{\psfig{figure=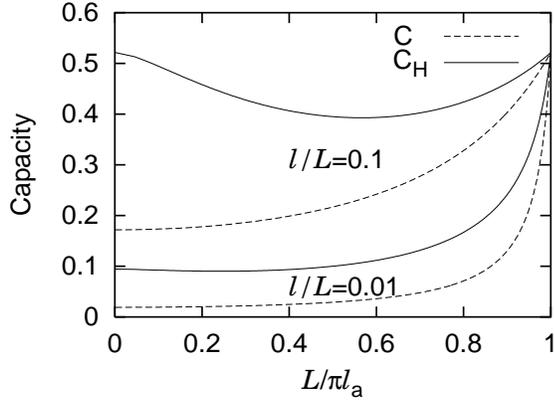,width=8cm}\medskip}
\caption[]
{Amplification dependence of the capacity $C$ for heterodyne
detection of coherent states [Eq.\ (\ref{result})] and the capacity
$C_{\rm H}$ for arbitrary detection [Eq.\ (\ref{averchi})].
The input power is fixed at $P/P_0N=1$ and two values
of $l/L$ are chosen.}
\label{holevo}
\end{figure}

As shown in Fig.\ \ref{holevo}, the dependence of $C_{\rm H}$ on the amount of
amplification is non-monotonic --- in contrast to the monotonically
increasing $C$. Weak amplification reduces the capacity $C_{\rm H}$,
while stronger amplification causes $C_{\rm H}$ to rise again to
the limit $C_\infty$ at the laser threshold. The initial decrease for
$l_a\gg L$ is described by
\begin{equation}
C_{\rm H}(L/l_a) \approx C_{\rm H}(0) -
(4 l L^2/3 l_a^2) \log_2 (\pi l_a/L).
\end{equation}

\begin{figure}
\centerline{\psfig{figure=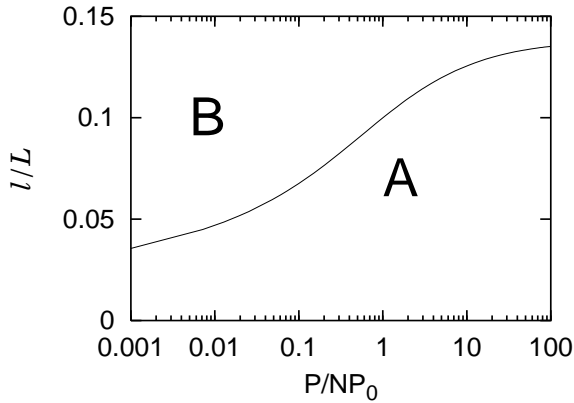,width=8cm}\medskip}
\caption[]
{Curve in parameter space separating region A
[in which $C_\infty > C_{\rm H}(0)$] from region B 
[in which $C_\infty < C_{\rm H}(0) $]. In region
A amplification of sufficient strength increases the capacity
$C_{\rm H}$, while in region B it does not.
}
\label{phase}
\end{figure}

Whether or not amplification ultimately increases $C_{\rm H}$
depends on the degree of disorder and on the input power.
We indicate by A the region in parameter space where
$C_\infty > C_{\rm H}(0) $ and by B the region
where $C_\infty < C_{\rm H}(0) $. In region A
strong amplification increases $C_{\rm H}$ while in region B
it does not. The separatrix is plotted in Fig.\ \ref{phase}.
For $P/N P_0 \ll 1$ 
the analytical expression for this curve separating regions A and B
is $P/N P_0= (3L/4l) \exp(-3L/8l+\gamma)$, while for 
$P/N P_0 \gg 1$ we find a saturation at
$l/L=3/8 e\approx 0.14$. This means that for $P/N P_0 \gg 1$ strong
amplification increases the capacity $C_{\rm H}$ provided
$l < 0.14 L$. 

At the laser threshold both $C$ and $C_{\rm H}$ reach the same
``universal'' limit $C_\infty$ given by Eq.\ (\ref{infty}), which
depends only on the dimensionless input power per mode
$P/NP_0$ and not on the degree of disorder. This remarkably 
rich interplay of multiple scattering and amplification is worth 
investigating experimentally, for example in the context of a 
random laser \cite{Wie97,Cao01}.

In conclusion, we have investigated the effect of amplification on the
information capacity of a disordered waveguide, focusing on the 
competing effects of stimulated and spontaneous emission. We have compared
the capacity $C$ for heterodyne detection of coherent states with
the Holevo bound $C_{\rm H}$ for arbitrary detection scheme. While
amplification increases $C$ for any magnitude of disorder and input
power, the effect on $C_{\rm H}$ can be either favorable or not, as is
illustrated by the ``phase diagram'' in Fig.\ \ref{phase}.

This research was supported by the Dutch Science Foundation
NWO/FOM and by the RTN program of the European Community.

\ecols
\end{document}